\documentclass[journal,onecolumn]{IEEEtran}
\usepackage{float}
\usepackage{textgreek}
\usepackage{amsmath}
\usepackage{graphicx}
\usepackage{tikz}  
\usepackage{pgfplots} 
\pgfplotsset{width=10cm,compat=1.9}  
\usepgfplotslibrary{fillbetween}
\usepackage{tkz-euclide} 
\usetikzlibrary{shapes.geometric, arrows} 
\usepackage{pgfplots} 
\usepackage{media9} 
\pgfplotsset{width=10cm,compat=1.9}  
\tikzstyle{startstop} = [rectangle, rounded corners, minimum width=3cm, minimum height=1cm,text centered, draw=black, fill=red!30]
\tikzstyle{repeat} = [rectangle, rounded corners, minimum width=3cm, minimum height=1cm,text centered, text width=3cm, draw=black, fill=yellow!30]
\tikzstyle{io} = [trapezium, trapezium left angle=70, trapezium right angle=110, minimum width=3cm, minimum height=1cm, text centered, draw=black, fill=blue!30]
\tikzstyle{process} = [rectangle, minimum width=3cm, minimum height=1cm, text centered, text width=3cm, draw=black, fill=orange!30]
\tikzstyle{decision} = [diamond, minimum width=3cm, minimum height=1cm, text centered, text width=3cm, draw=black, fill=green!30]
\tikzstyle{arrow} = [thick,->,>=stealth]

\begin{document}
%
\title{Measuring the Sun’s Core with Neutrino Measurements: A Solar Orbiter Concept}

\author{Jonathan Folkerts}

\markboth{Presented at 245th AAS meeting, National Harbor, MD, Jan 2025}{Presented at 245th AAS meeting, National Harbor, MD, Jan 2025}

\maketitle

\begin{abstract}
Traditional neutrino detectors are built deep underground to reduce backgrounds. The neutrino solar orbiting laboratory (\textnu{}SOL) collaboration has been developing a concept to improve neutrino measurement not with a larger detector underground, but instead we use the nuclear excitation from the neutrino interaction to produce a multi-pulse signal. Cerium-doped gadolinium aluminum gallium garnet (GAGG) is a new scintillator which has 23\% gallium by mass. When a neutrino interacts with the GAGG, about 10\% of the time it will be in an excited nuclear state rather than in the base energy level. A segmented detector looking for the pulses separated by distance and time has the potential to greatly limit background noise from solar wind, cosmic rays, and galactic gamma rays. A polar LEO CubeSat mission is currently in development to measure the GCR backgrounds outside the Van Allen Belts.

In this summary of my presentation I will quickly lay the groundwork of the interaction of interest and what a solar orbiter’s detector could look like. I will then explore what measurements a near-solar orbiter could make. With these measurements in mind, I will discuss the feasibility of a direct observation of the core’s shape, and I will discuss how a solar orbiter’s measurements could improve a Standard Solar Model search and compare that measurement with the current global neutrino measurements. I will conclude with a discussion of what these observables could tell us about the solar interior. 

\end{abstract}

\section{Solar Neutrino Measurement}
\IEEEPARstart{T}{he} Neutrino Solar Orbiting Laboratory project, \textnu{}SOL, has been pursuing the goal of designing and operating a neutrino detector capable of operating in a near-solar orbit. Solar neutrinos have several key scientific advantages. They escape the solar core because of their low interaction cross section, allowing for direct measurements of the core, the interaction always begins in a detector, allowing for an active veto volume to reject cosmic rays and solar wind, and it should be possible to build a neutrino detector capable of operating in space. The key disadvantages of solar neutrinos are that they are difficult to detect due to their small interaction cross section and that the earth is too large of a baseline for any sort of direct imaging.

Historically solar neutrinos have been measured using radiochemical means such as chlorine in the famous homestake mine experiment, or gallium in several followups. These experiments operated by chemically separating the resulting {}$^{71}$Ge or {}$^{37}$Ar and counting the number of final-state particles. More modern experiments used water cherenkov interactions to detect neutrinos. Our group and one other are looking into designing an online gallium experiment.

\begin{align}\label{eq:interaction1}
     ^{71}\text{Ga} + \nu_e&\to {}^{71}\text{Ge}^{*+} + e^-\\
   \label{eq:interaction2}{}^{71}\text{Ge}^{*+} &\to {}^{71}\text{Ge}^{+} + \gamma\text{(s)}
\end{align}

\begin{figure}[htbp]
    \centering
    \includegraphics[width=0.35\textwidth]{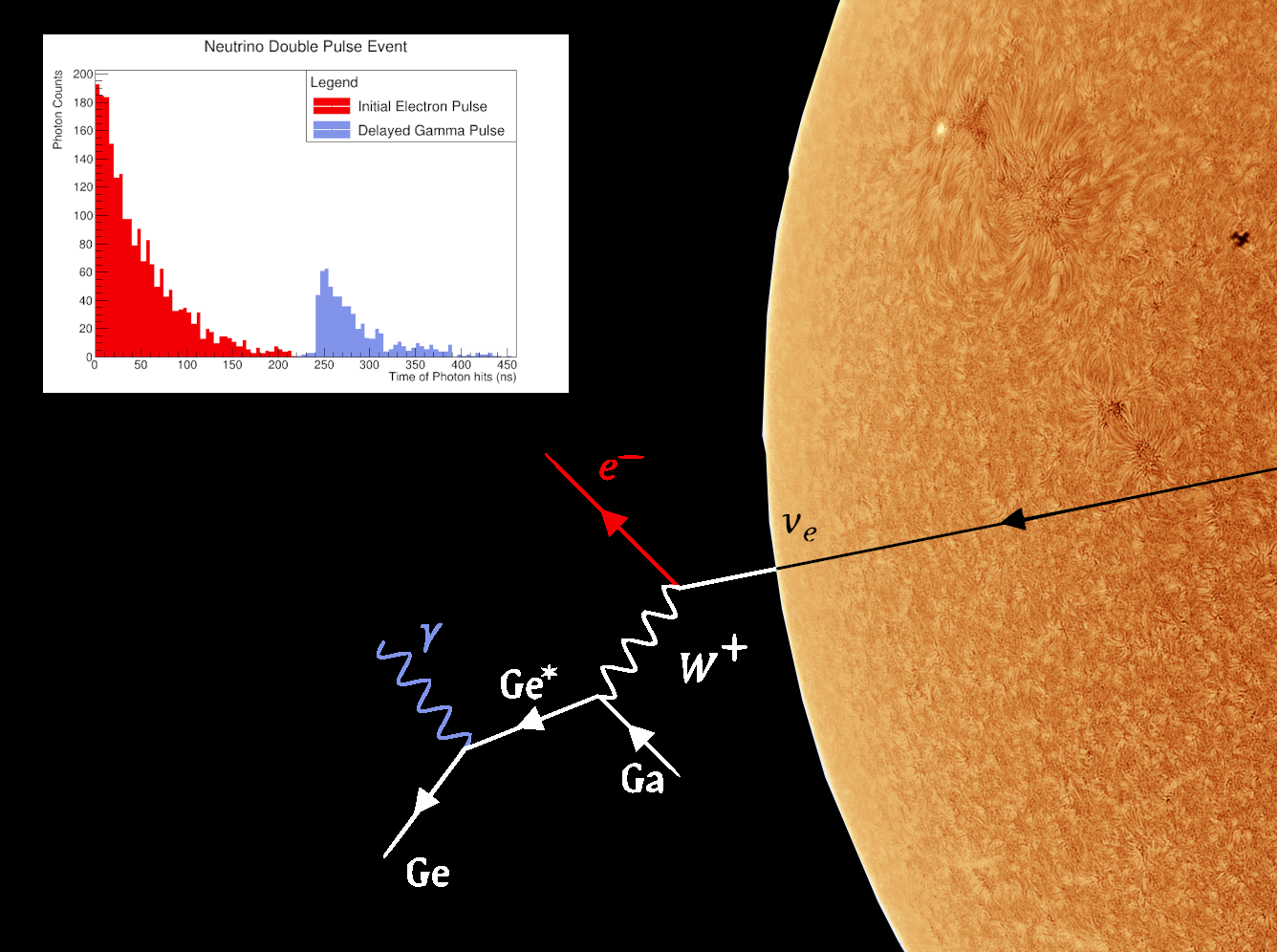}
    \caption{Diagram of gallium interaction with a solar neutrino resulting in a prompt electron and delayed gamma ray.}
    \label{fig:interactionDiagram}
\end{figure}

The interaction of interest is given in (\ref{eq:interaction1}) \& (\ref{eq:interaction2}), and it is also shown in Figure \ref{fig:interactionDiagram}. In this interaction, a solar neutrino interacts with a gallium nucleus producing an electron and a germanium nucleus in an excited state. This excited germanium then decays to ground, releasing one or more gamma rays. The excited state occurs in $\sim10\%$ of interactions, and $\sim50\%$ of those travel through the first excited state with characteristic energy 175 keV and 80 ns half life.

\begin{figure}[htbp]
    \centering
    \includegraphics[width=0.5\textwidth]{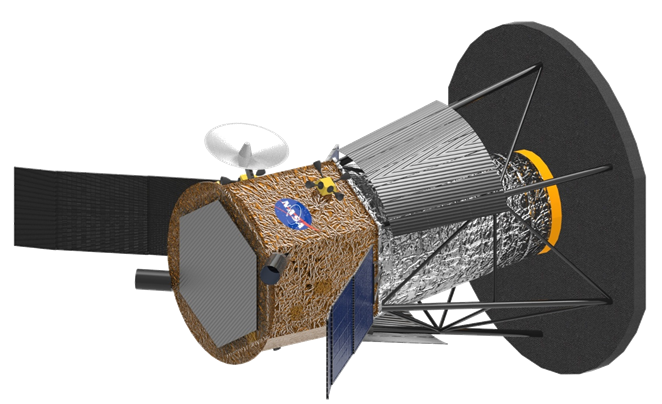}
    \caption{Possible design for the Neutrino Solar Orbiting Lab, produced by the MSFC Advanced Concept Office.}
    \label{fig:MSFCDesign}
\end{figure}

We envision an experiment flying on a spacecraft with a solar shield similar to the one used by Parker, or even the spare from Parker's construction. This would protect the electronics from the heat, and allow the spacecraft to operate close to the sun. An advanced concept office study at Marshal Space Flight Center created a possible design, shown in Figure \ref{fig:MSFCDesign}. If the detector were able to approach $7R_\odot$, the neutrino flux would be increased one thousandfold, and such an orbit seems possible.

The detector on-board this spacecraft would likely be constructed of a recently developed scintillator which is $\sim20\%$ gallium by mass, Gadolinium Aluminum Gallium Garnet (GAGG). This crystal is very fast with decay times between 50 and 100 ns. It is also very bright with a photon yield of 40-60 photons/keV. Because it is a scintillator, this detector would double as a charged-particle detector, and the gadolinium allows for it to be operated as a neutron detector as well. Another candidate scintillator is \textbeta-Ga$_2$O$_3$. This crystal is faster, $\sim70\%$ gallium, and still about 10\% the brightness of GAGG, but it is currently in its infancy as a scintillator.

\section{Possible Measurements}
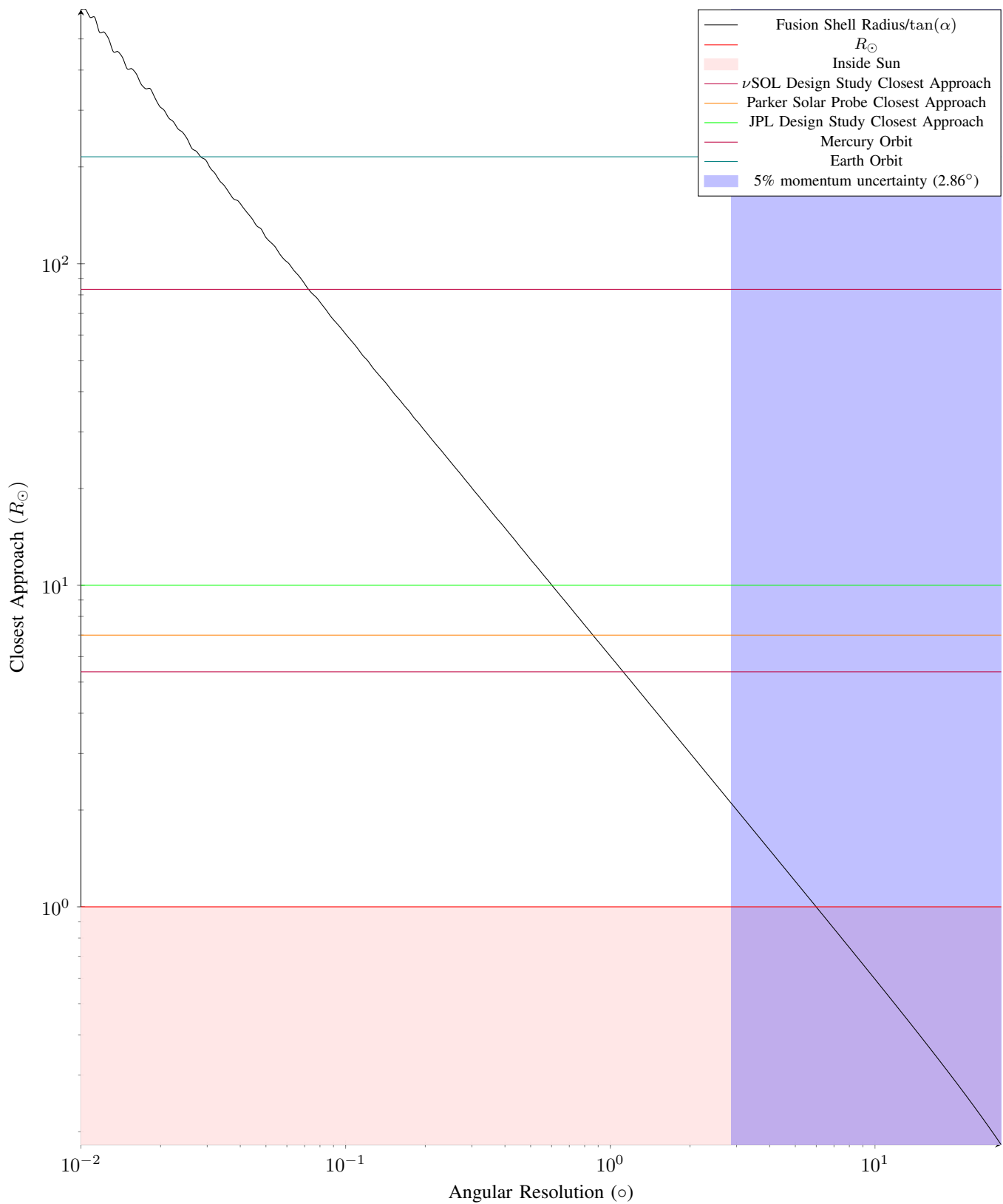
\begin{figure}[htbp]
  \centering
  \begin{tikzpicture}
\begin{loglogaxis}[
    axis lines = left,
    width=\textwidth,
    height=0.9\textheight,
    title = Required Closest Approach vs Desired Angular Resolution,
    xlabel = {Angular Resolution ($\circ$)},
    ylabel = {Closest Approach $(R_\odot)$},
    legend style={
        at={(1,1)}, 
        font=\footnotesize, 
        font=\fontsize{8pt}{9pt}\selectfont, 
        row sep=-1pt 
                },
]
\addplot [
    smooth,
    domain=1e-2:30, 
    samples = 200, 
    color=black,
]
{0.105/tan(x)};
\addlegendentry{Fusion Shell Radius/$\tan(\alpha)$}
\addplot [
        name path=f,
        smooth,
    domain=1e-2:30, 
        samples=10, 
        color=red,
    ]
    {1};
    \addlegendentry{$R_\odot$}
    \path[name path=axis] (axis cs:1e-5,1e-1) -- (axis cs:30,1e-1);
    \addplot[red!10, opacity=0.95] fill between[of=axis and f];
    \addlegendentry{Inside Sun}
    \addplot [
        name path=f,
        smooth,
    domain=1e-2:30, 
        samples=10, 
        color=purple,
    ]
    {5.3821020};
    \addlegendentry{$\nu$SOL Design Study Closest Approach}
        \addplot [
        name path=f,
        smooth,
    domain=1e-2:30, 
        samples=10, 
        color=orange,
    ]
    {7};
    \addlegendentry{Parker Solar Probe Closest Approach}
    \addplot [
        name path=f,
        smooth,
    domain=1e-2:30, 
        samples=10, 
        color=green,
    ]
    {10};
    \addlegendentry{JPL Design Study Closest Approach}
    \addplot [
        name path=f,
        smooth,
    domain=1e-2:30, 
        samples=10, 
        color=purple,
    ]
    {83.24};
    \addlegendentry{Mercury Orbit}
    \addplot [
        name path=f,
        smooth,
    domain=1e-2:30, 
        samples=10, 
        color=teal,
    ]
    {215};
    \addlegendentry{Earth Orbit}
    \path[name path=nova] (axis cs:2.86,1e-2) -- (axis cs:2.86,215e2);
    \path[name path=tall] (axis cs:30,1e-2) -- (axis cs:30,215e2);
    \addplot[blue, opacity=0.25] fill between[of=tall and nova];
    \addlegendentry{5\% momentum uncertainty (2.86$^\circ$)}
\end{loglogaxis}
\end{tikzpicture}
    \caption{Required closest approach for a spacecraft vs the angular resolution necessary to distinguish the top (left) of the fusion toroid from the bottom (right) is shown in black. The distance to the sun for several objects are plotted as the horizontal lines. A 5\% momentum uncertainty, typical for high-energy beamline neutrinos, is show in the blue region.}
    \label{fig:closest}
\end{figure}

For a detector to make a direct measurement of the solar interior, the minimum angular resolution to make a four-pixel image, i.e. an image that can distinguish the top of the fusion toroid from the bottom and the left of the fusion toroid from the right, is shown in Figure \ref{fig:closest}. In this figure, the angular resolution typical of high-energy beamline experiments is shown in the blue region. This resolution is much higher than typical for charged current solar neutrino experiments, which have nearly isotropic electrons in their final states. This implies that a direct measurement of the solar interior, even on an orbiter, is technically infeasible.

This does not mean that there is not science that can be done on a solar orbiter. Table \ref{tab:Orbits} shows the number of neutrinos that would be expected after a ten year mission for a potential orbit that end closer to the sun that Parker presently operates, and the global neutrino fit. For each case, the expected neutrino count is given with $1\sigma$ Poisson statistical uncertainties. The two spacecraft expectation counts are the rates if the excited state is required or if the ground state is allowed in the detector. To study what the neutrino measurement can do, we turn to solar modeling.

\begin{table}[htbp]
    \centering
    \begin{tabular}{|l|p{0.85in}|p{1.30in}|p{1.1in}|}
         \hline Detector & Duration & Expected Neutrino Count /100 kg Ga & Corresponding Luminosity $(L_\odot)$ \\\hline\hline
         $\nu$SOL All & 10 years & $149.7^{+12.3}_{-12.2}$ & $1^{+8.22\%}_{-8.17\%}$\\
         Transitions&&&\\\hline
         $\nu$SOL Excited  & 10 years & $ 15.7^{+4.2}_{-3.9}$ & $1^{+26.6\%}_{-25.0\%}$\\
         Transitions& & &\\\hline
         Borexino & Combined data \cite{ref:borexinoLuminosityMeasurement} & - & $1.038^{+6.6\%}_{-5.8\%}$\\\hline
    \end{tabular}
    \caption{Table of the possible measurements of the $\nu$SOL spacecraft and the Borexino combined data for comparison.}
    \label{tab:Orbits}
\end{table}

\section{Solar Modeling}

A standard solar model (SSM) begins as a 1 $M_\odot$ star starting at either pre-main sequence cloud of gas or as a zero-age main sequence (ZAMS) star. In the model I am using, I initially begin with the pre-main sequence star, but for computational saving, the star model is loaded from its ZAMS model, modified by the searching function, and runs until hitting the present day solar age, $\tau_\odot$. This model assumes a very short pre-ZAMS time and no appreciable mass loss from ZAMS to present.

When a SSM has fully evolved, it is required to match three present-day parameters. It must match the present solar luminosity, $L_\odot$, the present solar radius, $R_\odot$, and the present surface metal-to-hydrogen mass fraction, $(Z/X)_\odot$. Of note, the least-well constrained of these parameters is $(Z/X)_\odot$, which has lead to several different models of metallicity which have been adopted over the years. An SSM will state which of these metallicity models it is using, such as  GS98 \cite{ref:MetallicityGS98}, which is a high-metallicity model, and AGS09 \cite{ref:MetallicityAGS09}, which is a low metallicity model. These names are derived from the authors of the paper introducing them, and the year of the paper. Recently, the Borexino collaboration has made measurements which disfavor low-metallicity solar models at 3.1$\sigma$, and so for my work here I will be focusing only on the GS98 metallicity model.

The three free parameters of the modeling are the initial fractions of helium and metals, $Y_\text{ini}$ and $Z_\text{ini}$, as well as the mixing-length parameter, $\alpha_\text{MLT}$ of mixing length theory. A SSM uses the minimal physical models needed to evolve a star in a way that matches nature. These processes include convective and radiative energy transport through the star, compositional changes driven by nuclear fusion, and microscopic diffusion of elements throughout the star\cite{ref:StandardSolarModelsReview}.

The tweak I make with solar neutrinos is to add an additional constraint inside the star. The neutrino flux provides a direct measurement of the power being generated during fusion inside the star's core.

\begin{figure}
        \centering
        \includegraphics[width=0.85\textwidth]{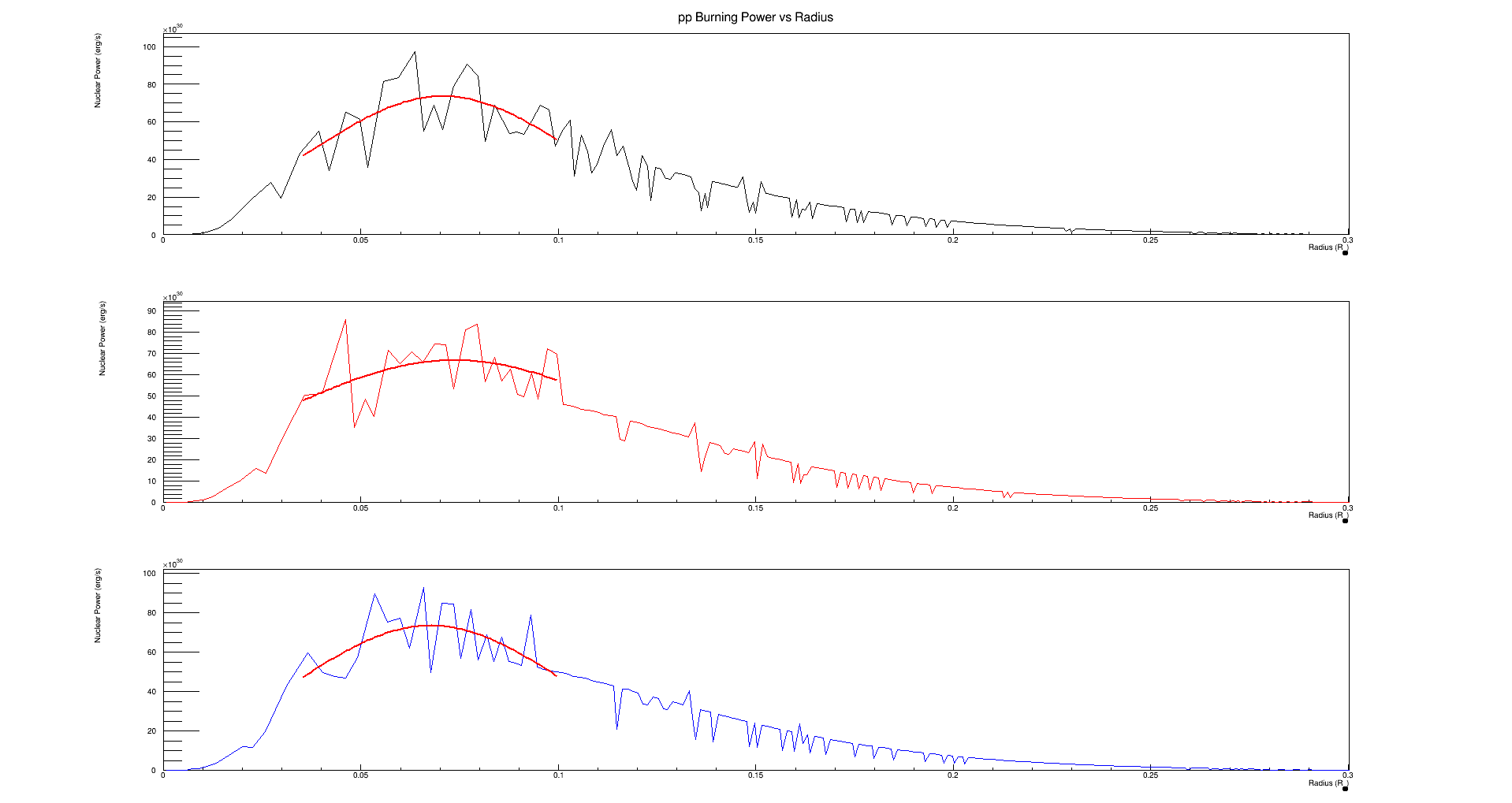}
        \caption{\textit{pp} Burning Power Profile vs Radius for the best model in a solar model simplex search for an SSM (top) the $\nu$SOL best fit (middle) and the global neutrino fit (bottom).}
        \label{fig:rawNucPP}
    \end{figure}

\begin{figure}
        \centering
        \includegraphics[width=0.85\textwidth]{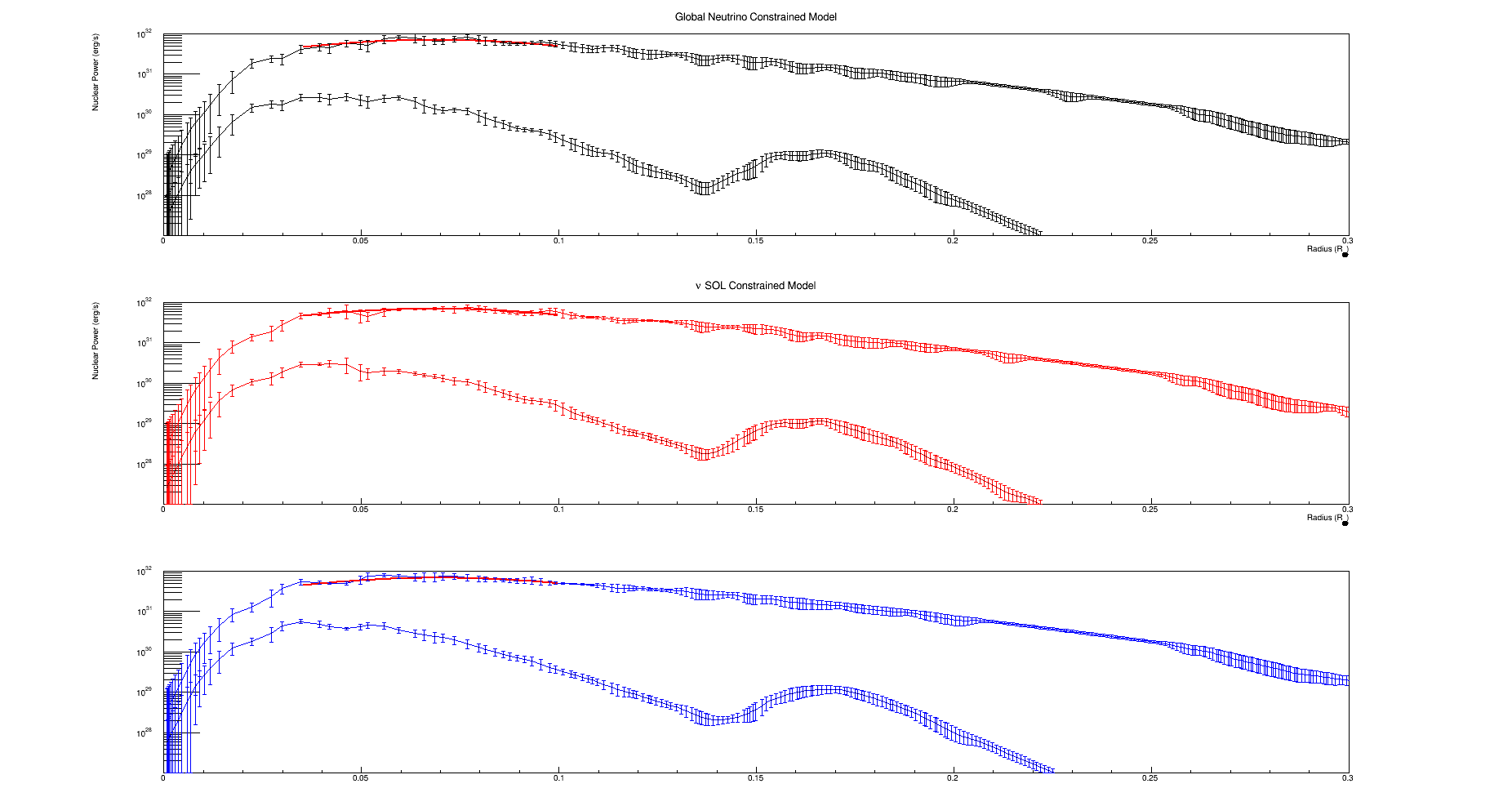}
        \caption{\textit{pp} and CNO Burning Power Profile vs Radius for the best model in a solar model simplex search for an SSM (top) the $\nu$SOL best fit (middle) and the global neutrino fit (bottom).}
        \label{fig:smoothedNucPP}
    \end{figure}

\begin{figure}
        \centering
        \includegraphics[width=0.85\textwidth]{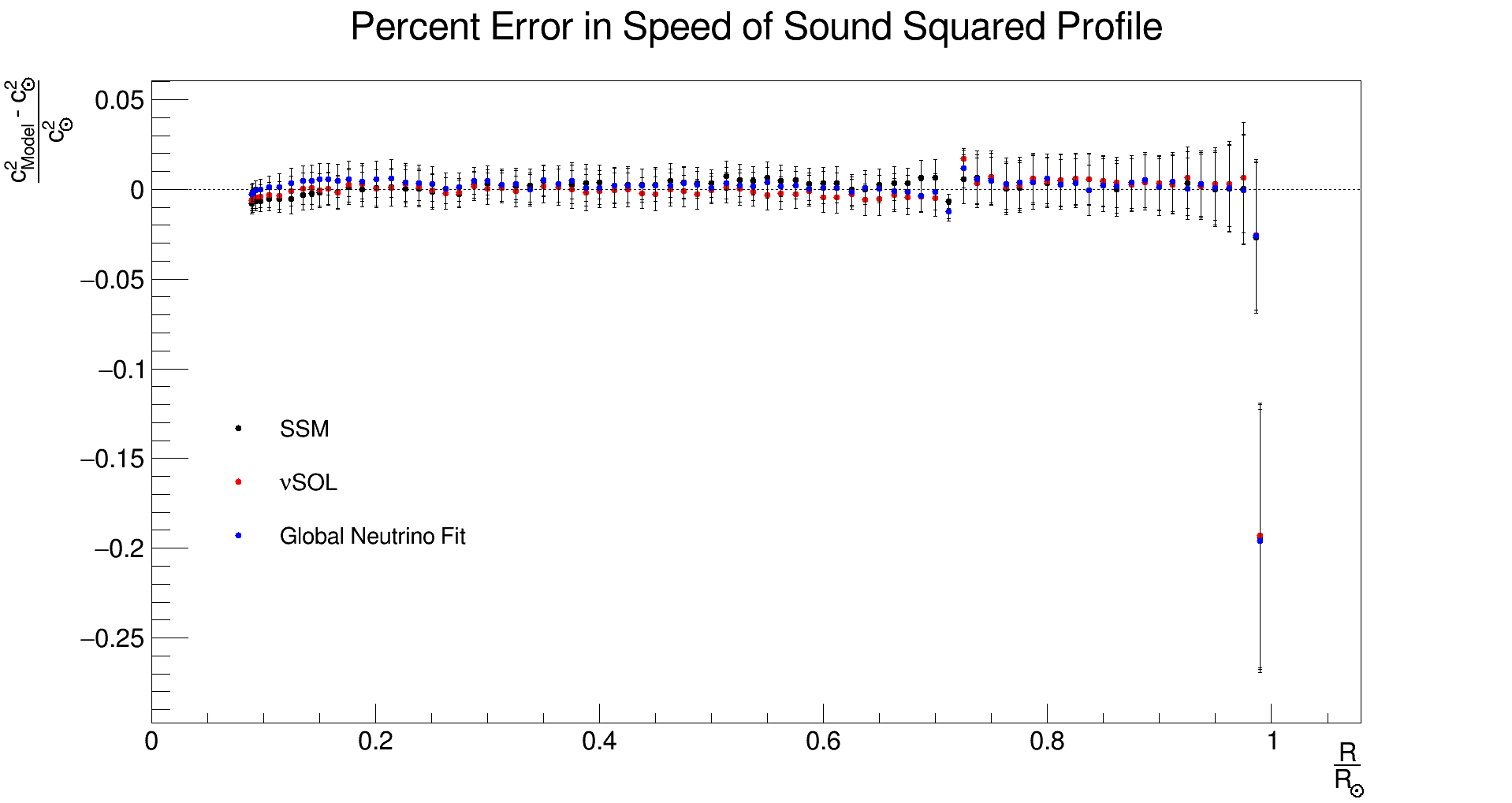}
        \caption{Percent error in the speed of sound profile for the solar models.}
        \label{fig:CSoundSquaredError}
    \end{figure}

The results of adding this constraint have a few interesting results. Figures \ref{fig:rawNucPP} and 5\ref{fig:smoothedNucPP} show the nuclear burning power profile for three simulations using an SSM, the $\nu$SOL constraint, and the global neutrino constraint. In each, the shape of the burning profile has changed slightly, and the centers of their peak burning is statistically different from one another. A common place that solar models go to the cutting room floor is in their speed of sound profile. As we can see in Figure \ref{fig:CSoundSquaredError}, all three models have very good agreement with the measured speed of sound profile from helioseismology.

\begin{figure}
        \centering
        \includegraphics[width=0.85\textwidth]{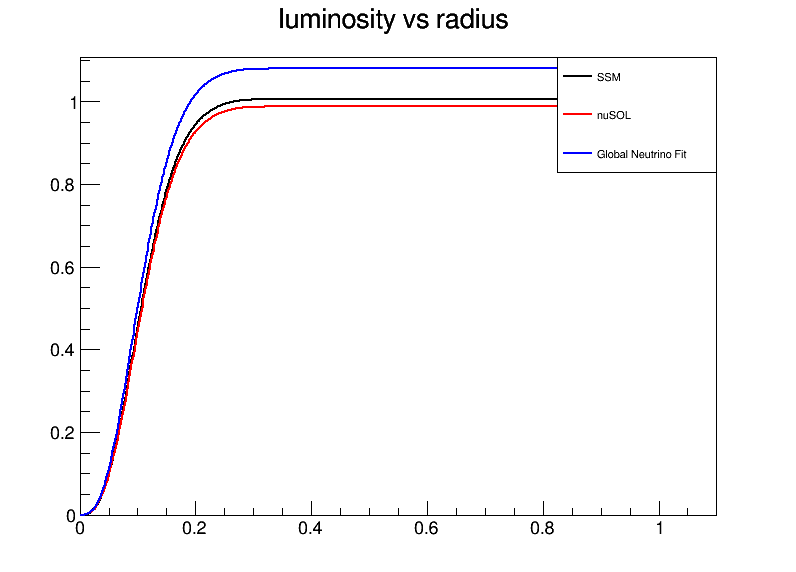}
        \caption{Solar luminosity profile vs radius for the solar models.}
        \label{fig:Luminosity}
    \end{figure}

One particularly interesting result from these simulations is the luminosity vs radius graph in Figure \ref{fig:Luminosity}. Unsurprisingly, a higher measured solar luminosity resulting from the global neutrino fit gives a higher simulated luminosity. What is interesting in these results is that adding a neutrino constraint with the $\nu$SOL possible measurement seems to slightly lower the final luminosity of the sun. Why this happens is unclear, but I am working currently to try and unravel this mystery.

\newpage
\bibliographystyle{ieeetr}
\bibliography{yourbibfile}

\end{document}